\documentclass[pre,twocolumn,reprint,superscriptaddress]{revtex4-1}
\usepackage{amsmath,amssymb,amsthm,bm,braket,ascmac,bbm}
\usepackage{graphicx}
\usepackage[usenames]{color}
\theoremstyle{definition}

\newcommand{\dd}{\mathrm{d}}
\newcommand{\ee}{\mathrm{e}}
\newcommand{\ii}{\mathrm{i}}

\newcommand{\G}{\mathrm{G}}
\newcommand{\Lc}{1}
\newcommand{\Rc}{2}
\newcommand{\kett}[1]{\ket{#1}\!\rangle}
\newcommand{\braa}[1]{\langle\!\bra{#1}}

\newcommand{\gs}{\text{GS};\alpha}
\newcommand{\hrho}{\hat{\rho}}
\newcommand{\hO}{\hat{O}}

\begin{document}
\title{
Entanglement Pre-thermalization
in an Interaction Quench between Two Harmonic Oscillators}

\author{Tatsuhiko N. Ikeda}
\affiliation{Institute for Solid State Physics, University of Tokyo, Kashiwa, Chiba 277-8581, Japan}
\affiliation{Department of Physics, Harvard University, Cambridge, Massachusetts 02138, USA}
\affiliation{Department of Physics, University of Tokyo, Bunkyo-ku, Tokyo 113-0033, Japan}
\author{Takashi Mori}
\affiliation{Department of Physics, University of Tokyo, Bunkyo-ku, Tokyo 113-0033, Japan}
\author{Eriko Kaminishi}
\affiliation{Department of Physics, University of Tokyo, Bunkyo-ku, Tokyo 113-0033, Japan}
\author{Masahito Ueda}
\affiliation{Department of Physics, University of Tokyo, Bunkyo-ku, Tokyo 113-0033, Japan}
\affiliation{RIKEN Center for Emergent Matter Science (CEMS), Wako, Saitama 351-0198, Japan}

\date{\today}
\begin{abstract}
Entanglement pre-thermalization (EP) is a quasi-stationary nonequilibrium state of a composite system
in which each individual subsystem looks thermal but the entire system remains nonthermal due to quantum entanglement between subsystems.
We theoretically study the dynamics of EP
following a coherent split of a one-dimensional harmonic potential in which two interacting bosons are confined.
This problem is equivalent to that of an interaction quench between two harmonic oscillators.
We show that this simple model captures the bare essentials of EP;
that is, each subsystem relaxes to an approximate thermal equilibrium,
whereas the total system remains entangled.
We find that a generalized Gibbs ensemble, which incorporates nonlocal conserved quantities, exactly describes the total system.
In the presence of a symmetry-breaking perturbation, EP is quasi-stationary and eventually reaches thermal equilibrium.
We analytically show that 
the lifetime of EP is inversely proportional to the magnitude of the perturbation.
\end{abstract}
\maketitle

\section{Introduction}
Over the last two decades or so,
the foundation of quantum statistical mechanics has seen a resurgence of interest~\cite{Polkovnikov2011c,Eisert2015,DAlessio2015,Gogolin2015a},
motivated in large part by the advances in
ultracold atom experiments~\cite{Kinoshita2006,Trotzky2011,Gring2012,Langen2013}.
One of the fundamental questions is how statistical mechanics emerges out of the unitary time evolution
of quantum mechanics~\cite{Neumann2010,Goldstein2010b,Reimann2008,Linden2009,Reimann2012,Tasaki2016}.
Recent studies have clarified that an isolated quantum system can effectively reach thermal equilibrium,
where the memory of the initial state is not lost but only hidden in many-body correlations which
cannot be detected with few-body observables~\cite{Nandkishore2015}.
On the other hand, several exceptions have been found where
stationary states are not thermal.
For example,
non-thermal steady states can be described by
the generalized Gibbs ensemble 
in systems that have a large number of conserved quantities
as represented by integrable systems~\cite{Cazalilla2006,Rigol2007,Kollar2008,Rigol2009,Calabrese2011,Cazalilla2012a,Caux2012,Fagotti2013,Pozsgay2013,Wouters2014,Langen2015,Hamazaki2016}.
A system exhibiting many-body localization~\cite{Basko2006,Pal2010a,Altman2015}
does not thermalize either, since
a random potential tends to localize particles, thereby producing a large number of effective conserved quantities.
Thus,
one is naturally led to the question concerning
under what conditions conventional statistical mechanics holds true
and what exceptional stationary states are possible.

Entanglement pre-thermalization (EP)~\cite{Kaminishi2015}
offers yet another exception and 
has recently been proposed in a situation in which
an isolated one-dimensional Bose gas
is coherently split into two.
Each individual subsystem subsequently relaxes to a stationary state
which can be well described by the canonical ensemble at an effective temperature,
whereas the stationary state of the total system cannot be described by the canonical ensemble at any temperature
due to quantum entanglement between the subsystems.
Thus, entanglement brings the system into a new class of prethermalized states.
However, 
it remains to be clarified
what statistical-mechanical ensemble is suited to describe EP.

In this paper, 
we address this question
by considering
a coherent split of a harmonic potential,
in which two identical bosons are confined.
This problem is shown to be equivalent to 
that of the interaction quench of two harmonic oscillators.
We show that this simple model captures the bare essentials of EP
and find that
a generalized Gibbs ensemble involving nonlocal conserved quantities describe EP exactly.
This implies that
EP can be understood if we take into account appropriate nonlocal conserved quantities
which retain quantum entanglement between the subsystems.
We also consider the effect of a symmetry-breaking perturbation that causes EP to relax into a true equilibrium state
and show that the lifetime of EP is inversely proportional to the magnitude of the perturbation.

The rest of this paper is organized as follows.
In Sec.~\ref{sec:setup},
we formulate our problem
and give analytical expressions for the density matrix that
describes the long-time average of physical quantities.
In Sec.~\ref{sec:key_features},
we demonstrate
that in the long-time average each subsystem is approximately thermal 
and that entanglement is retained within degenerate subspaces.
Thus the essential feature of EP is captured by our model.
In Sec.~\ref{sec:ensembles},
we show that the long-time average can be described well by
neither the canonical ensemble nor its simple generalization,
and that it is described exactly by the generalized Gibbs ensemble
that incorporates nonlocal conserved quantities.
In Sec.~\ref{sec:plateau},
we consider the effects of a symmetry-breaking perturbation
and show that EP exhibits a plateau that eventually relaxes to a true equilibrium state
due to the symmetry-breaking perturbation.
In Sec.~\ref{sec:conclusions},
we summarize our results and
discuss some future directions.

\begin{figure}
\begin{center}
\includegraphics[width=7cm]{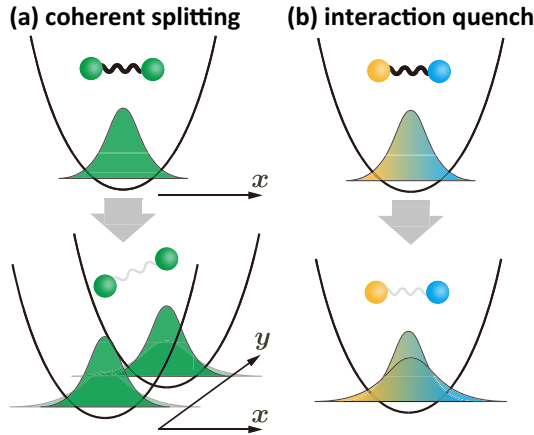}
\caption{
(Color Online)
(a) Two interacting identical bosons are initially confined in a harmonic potential along the $x$ axis.
Then the gas is coherently split into two in the $y$ direction.
After the split,
we post-select only those states which involve
one boson in each potential.
(b) The interaction between two distinguishable particles is suddenly switched off.
The subsequent time evolution
and the final state are equivalent to those of case (a).
}
\label{fig:schematic}
\end{center}
\end{figure}

\section{Setup}\label{sec:setup}
We consider a situation in which two identical bosons are confined in a one-dimensional harmonic potential
and they interact with each other via a quadratic interaction $V(x_1,x_2)=\alpha^2(x_1-x_2)^2/2$,
where $\alpha(>0)$ denotes the strength of the interaction.
We suppose that these two bosons are initially in the ground state,
which is denoted in the second-quantized form as
$\int \dd x_1 \dd x_2 \Psi_G(x_1,x_2)\psi^\dag(x_1)\psi^\dag(x_2)\ket{\Omega}$,
where $\psi(x)$ is the bosonic field operator, $\ket{\Omega}$ is the vacuum state,
and $\Psi_\G(x_1,x_2)$ $(=\Psi_\G(x_2,x_1))$ is the two-body wave function in the ground state.

At time $t=0$, the harmonic potential is split into two identical copies, 1 and 2,
as illustrated in Fig.~\ref{fig:schematic}(a).
In this process,
each boson is assumed to be converted to the superposition of two states,
one in potential 1 and the other in potential 2.
Thus, the state just after the split is given by the replacement $\psi(x)\to [\psi_\Lc(x)+\psi_\Rc(x)]/\sqrt{2}$,
where $\psi_\Lc(x)$ and $\psi_\Rc(x)$ are the bosonic field operators in potentials 1 and 2, respectively, that commute with each other.
The resulting state involves three cases: (i) two bosons are in potential 1,
(ii) two bosons are in potential 2,
and (iii) one boson lies in each potential.
We project out the cases (i) and (ii) since they do not contribute to the nonlocal correlation
such as $\psi^\dag_\Lc(x_1) \psi_\Rc(x_2)$,
where $x_i$ ($i=1,2$) refers to the coordinate in potential $i$. 
Thus, we post-select the state
\begin{align}\label{eq:initial}
\ket{\psi_0}=\int \dd x_1 \dd x_2 \Psi_\G (x_1,x_2) \psi_\Lc^\dag (x_1) \psi_\Rc^\dag (x_2) \ket{\Omega}
\end{align}
as the one after the split.
For $t>0$, the system evolves from the initial state~\eqref{eq:initial}
as two independent harmonic oscillators
since we assume that the two potentials are sufficiently displaced from each other so that
the two bosons have no overlap and hence do not interact with each other.

The coherent splitting thus defined is equivalent to an interaction quench
between two distinguishable particles with equal mass
because quantum statistics plays no role after the split.
(Note that the zero-temperature ground state of bosons is the same as that of distinguishable particles having the same mass.)
We therefore consider the following Hamiltonian
\begin{align}\label{eq:ham}
\hat{H}_\alpha= 
\frac{1}{2}(\hat{p}_1^2+\hat{p}_2^2)
+\frac{1}{2}(\hat{x}_1^2+\hat{x}_2^2)
+\frac{\alpha^2}{2}(\hat{x}_1-\hat{x}_2)^2,
\end{align}
where the mass $m$ of the particles and the frequency $\omega$ of the harmonic potential are set to unity,
and the dynamical variables $\hat{x}_i$ and $\hat{p}_i$ $(i=1,2)$ are assumed to satisfy the canonical commutation relations.
Then the ground state $\ket{{\rm GS};\alpha}$ of Eq.~\eqref{eq:ham} takes the form of Eq.~\eqref{eq:initial}.
At time $t=0$, the strength of interaction $\alpha$ is suddenly changed to zero,
and the system evolves in time according to $\hat{H}_{\alpha=0}$.
In the following, we use the description of the interaction quench rather than
that of the coherent splitting.

The Hamiltonian~\eqref{eq:ham} is easily diagonalized
for $\alpha=0$
since it represents two noninteracting harmonic oscillators.
We define the annihilation operators for each particle
$a_i \equiv\frac{1}{\sqrt{2}}(x_i +\ii p_i)$ $(i=1,2)$,
and their linear combinations $a_\pm \equiv \frac{1}{\sqrt{2}}(a_1\pm a_2)$.
Then we obtain the diagonalized Hamiltonian
\begin{align}\label{eq:ham0}
\hat{H}_{\alpha=0}=a_1^\dag a_1+a_2^\dag a_2 =a_+^\dag a_+ + a_-^\dag a_-,
\end{align}
where we have omitted the zero-point energy.
We denote by $\ket{0}$ the Fock vacuum defined by $a_1\ket{0}=a_2\ket{0}=0$.
Then, the energy eigenstates are represented in two ways as
\begin{align}
\ket{m,n} = \frac{(a_1^\dag)^m}{\sqrt{m!}}\frac{(a_2^\dag)^n}{\sqrt{n!}}\ket{0},\
\kett{m,n} = \frac{(a_+^\dag)^m}{\sqrt{m!}}\frac{(a_-^\dag)^n}{\sqrt{n!}}\ket{0}.\label{eq:pm_basis}
\end{align}
We note that the degenerate subspace with eigenenergy $N$ (in units of $\hbar\omega$) is given by
\begin{align}\label{eq:degen_sub}
\mathcal{H}_N \equiv \text{span} \{ \ket{m,n} | m+n=N\},
\end{align}
which can also be defined in terms of $\kett{m,n}$.

For $\alpha\neq0$,
the Hamiltonian~\eqref{eq:ham} is diagonalized
through the changes of variables
corresponding to the center-of-mass motion and the relative motion.
Namely, we introduce  $a_{\rm CM} = a_+$ and 
$a_{\rm rel} = \cosh r a_- + \sinh r a_-^\dag$
with
\begin{align}
\ee^{4r}=1+\alpha^2.
\end{align}
Then we obtain
$\hat{H}_\alpha = a_{\rm CM}^\dag a_{\rm CM} +\sqrt{1+\alpha^2} a_{\rm rel}^\dag a_{\rm rel}$.
The ground state $\ket{{\rm GS};\alpha}$ of $\hat{H}_\alpha$ is defined by the following conditions:
$a_{\rm CM}\ket{{\rm GS};\alpha} = a_{\rm rel}\ket{{\rm GS};\alpha}=0$.
We note that $\ket{{\rm GS};\alpha}$
is a squeezed vacuum:
\begin{align}\label{eq:GSalpha}
\ket{{\rm GS};\alpha}
=\frac{1}{\sqrt{\cosh r}}\ee^{-\frac{\tanh r}{2}(a_-^\dag)^2}\ket{0}.
\end{align}

The time evolution is obtained as
\begin{align}\label{eq:timet}
\ket{\Psi(t)} = \ee^{-\ii \hat{H}_{\alpha=0} t} \ket{{\rm GS};\alpha}
=\sum_{N=0}^\infty (-1)^N\sqrt{q_N} \ee^{-2\ii N t} \ket{\Phi_N},
\end{align}
where $\ket{\Phi_N}=\kett{0,2N}$ and
\begin{align}\label{eq:def_qN}
q_N=\frac{1}{\cosh r} \frac{(2N)!}{(N!)^2}\left( \frac{\tanh r}{2}\right)^{2N}.
\end{align}
In the following, we focus on
the long-time average, which is described by the density matrix $\rho_\infty$ defined by
\begin{align}\label{eq:lta_rho}
\rho_\infty = \overline{\ket{\Psi(t)}\bra{\Psi(t)}}=\sum_{N=0}^\infty q_N \ket{\Phi_N}\bra{\Phi_N},
\end{align}
where $\overline{f(t)}\equiv \lim_{T\to\infty}T^{-1}\int_0^T \frac \dd t f(t)$.
We note that, since the energy spectrum has equal spacings,
the temporal fluctuation around the long-time average is not negligible.
In more realistic interacting models, however,
this fluctuation is suppressed 
and the long-time average~\eqref{eq:lta_rho} is known to describe
an effective stationary state~\cite{Reimann2008,Short2011}.
Thus, we ignore the persistent temporal fluctuation in our model,
and assume that the long-time average in our model
gives an equivalent of the effective stationary state in realistic models.

\section{Key features of Entanglement Pre-thermalization}\label{sec:key_features}
In this section,
we show that our simple model captures the key features of EP.
Namely,
in the long-time average~\eqref{eq:lta_rho},
each subsystem approximately reaches thermal equilibrium,
and the diagonal/off-diagonal decomposition
holds true for the total system exactly as in Ref.~\cite{Kaminishi2015},
where all the information about subsystems is contained in the diagonal part
and EP is characterized by the off-diagonal part.

\subsection{Asymptotically thermal subsystem}
\begin{figure}
\begin{center}
\includegraphics[width=7cm,angle=0]{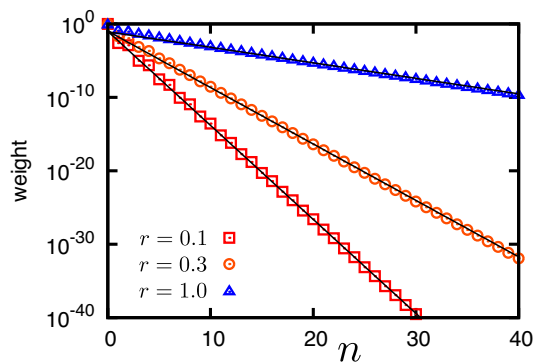}
\caption{(Color Online)
Weight $w_n$ of each energy eigenstate of the subsystem
for $r=0.1$ (square), 0.3 (circle), and 1.0 (triangle). 
The solid line shows the canonical weight $A\ee^{-\beta n}$
at the inverse temperature~\eqref{eq:temp},
where $A$ is determined by the least-squares fitting.
}
\label{fig:weight}
\end{center}
\end{figure}

To show the first key feature of EP that
each subsystem can be described by a conventional statistical ensemble,
let us consider the reduced density matrix of the subsystem 1:
\begin{align}
\rho^{(1)}_\infty = \text{tr}_2 \rho_\infty,
\end{align}
where $\text{tr}_{2}$ denotes the trace over the Hilbert space of the subsystem 2.
Due to symmetry under the interchange of 1 and 2, the following argument also applies to $\rho^{(2)}_\infty = \text{tr}_1 \rho_\infty$.
To take the partial trace,
we expand $\ket{\Phi_N}$ in terms of $\ket{m,n}$ as $\ket{\Phi_N}=\sum_{m=0}^{2N}c_{m,2N-m}\ket{m,2N-m}$
with $c_{m,n}\equiv\binom{m+n}{m}(-1)^m \sqrt{m!n!}$,
which leads to
\begin{align}\label{eq:phiphi}
\ket{\Phi_N}\bra{\Phi_N}
&=\sum_{m=0}^{2N}\sum_{m'=0}^{2N}c_{m,2N-m}c_{m',2N-m'}^*\notag\\
&\qquad\times\ket{m,2N-m}\bra{m',2N-m'}.
\end{align}
From Eqs.~\eqref{eq:lta_rho} and \eqref{eq:phiphi}, we obtain
\begin{align}\label{eq:lta_sub}
\rho^{(1)}_\infty = \sum_{m=0}^\infty w_m \ket{m}\bra{m},
\end{align}
where $w_m=\sum_{N= \lfloor m/2 \rfloor}^\infty q_N|c_{m,2N-m}|^2$
with $q_N$ defined in Eq.~\eqref{eq:def_qN}
represents the weight of each energy eigenstate
and can be represented in terms of the hypergeometric function.
By looking into the asymptotic behavior of $w_n$,
we obtain $w_n \propto \ee^{-\beta n}$ for $n\gg1$ with
\begin{align}\label{eq:temp}
\beta \equiv \ln \left( 2\coth r-1\right),
\end{align}
which means that Eq.~\eqref{eq:lta_sub} is asymptotically thermal.
The comparison between the actual weight $w_n$ 
and the asymptotic behavior is illustrated in Fig.~\ref{fig:weight}.
The approach to the asymptotic form is quite fast.

We note that $\beta$ can be regarded as the real temperature of the subsystem
that is calculated from the total energy
because the reduced density matrix $\rho^{(1)}_\infty$ itself is approximately thermal.
In contrast, in Ref.~\cite{Kaminishi2015},
the reduced density matrix is not necessarily close to a thermal one
and the temperature of the subsystem is determined by the least-squares fitting of correlation functions.

We remark that our initial state is different from the so-called two-mode squeezed state (TMSS),
which is given by $\ket{{\rm TMSS}}\propto \sum_{N=0} \lambda^N \ket{N,N}$ for $|\lambda|<1$.
The TMSS reduces to an exact thermal equilibrium state when either subsystem is traced out.
However, this is not appropriate for addressing EP
because, in each degenerate subspace $\mathcal{H}_{2N}$,
the state $\ket{N,N}$ is a direct product of states in the subsystems,
and therefore cannot support EP
which requires entanglement, namely, a superposition state between degenerate subspaces.

\subsection{Diagonal/off-diagonal decomposition}
Here we show the second key feature of EP that
the total system in the long-time average $\rho_\infty$ retains
entanglement within degenerate subspaces.
To this end, we show that
the diagonal/off-diagonal decomposition
\begin{align}
\rho_\infty=
\rho_\infty^\text{d}
+\rho_\infty^\text{off-d}
\end{align}
works exactly as in Ref.~\cite{Kaminishi2015}.
Here the diagonal $\rho_\infty^\text{d}$ and off-diagonal $\rho_\infty^\text{off-d}$ parts are given, respectively, as
\begin{align}\label{eq:diagonal}
\rho_\infty^\text{d} = \sum_{N=0}^\infty q_N \sum_{m=0}^{2N} |c_{m,2N-m}|^2 \ket{m,2N-m}\bra{m,2N-m}
\end{align}
and
\begin{align}\label{eq:offdiagonal}
\rho_\infty^\text{off-d} &= \sum_{N=0}^\infty q_N \sum_{m=0}^{2N}\sum_{\substack{m'=0\\ (m'\neq m)}}^{2N} c_{m,2N-m}c_{m',2N-m'}^*\notag\\
&\qquad\qquad\times \ket{m,2N-m}\bra{m',2N-m'}.
\end{align}

The off-diagonal part has no physical effect as long as we look at each subsystem alone.
In fact, it vanishes when either subsystem is traced out,
$\text{tr}_i \rho_\infty^\text{off-d} =0$ $(i=1,2)$.
Therefore, the diagonal part contains all the information about individual subsystems
in the sense that 
\begin{align}
\text{tr}_i\rho_\infty^\text{d}=\text{tr}_i\rho_\infty \quad (i=1,2).
\end{align}

\begin{figure}
\begin{center}
\includegraphics[width=8.5cm]{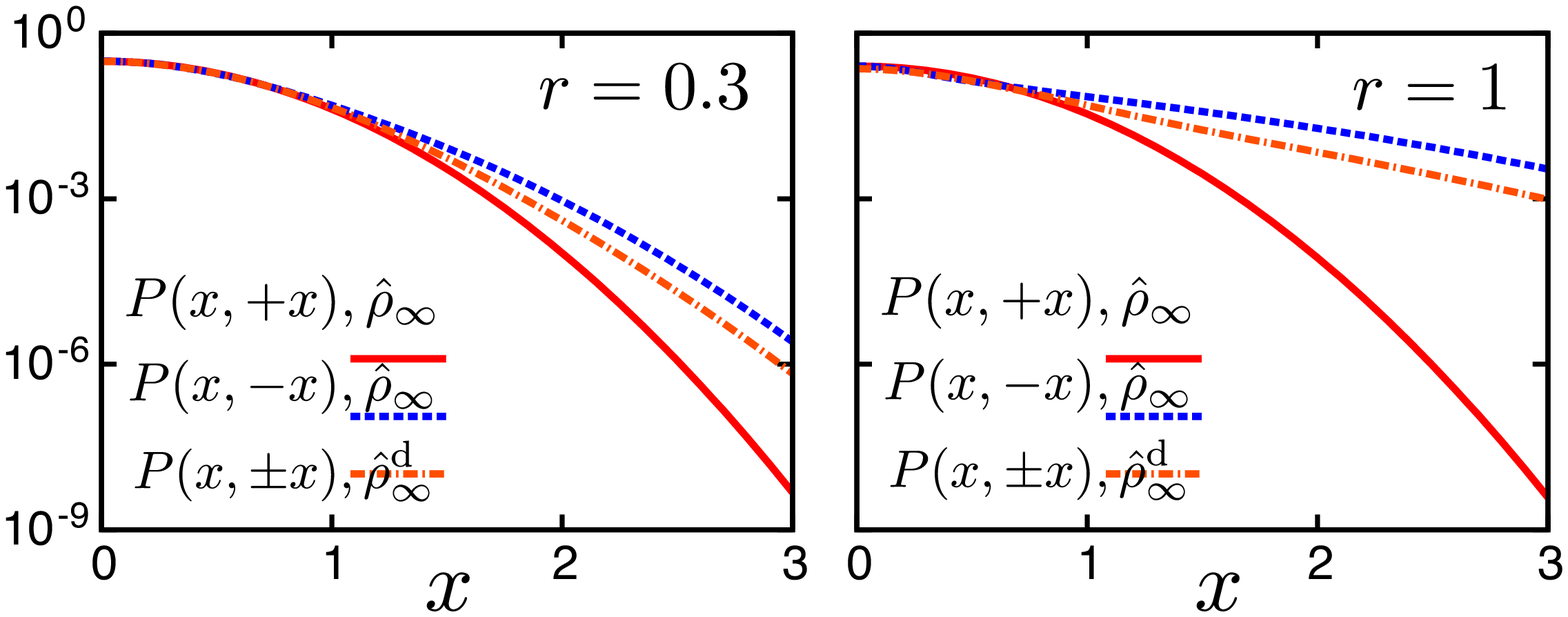}
\caption{(Color Online)
Distribution $P(x,y)$ with $x=y$ (solid),
$x=-y$ (dotted) calculated for the long-time average~\eqref{eq:lta_rho},
and $x=\pm y$ calculated for the diagonal part~\eqref{eq:diagonal} (dash-dotted).
Note that in the last case $P(x,x)=P(x,-x)$.
They are calculated for the quench magnitudes of $r=0.3$ (left) and 1 (right).
}
\label{fig:diagonal}
\end{center}
\end{figure}

Nonetheless,
the off-diagonal part plays an essential role in the nonlocal correlation
by which we mean the correlation between the different subsystems.
To show this, we investigate the joint probability distribution in the coordinate basis
\begin{align}\label{eq:joint}
P(x,y) = \braket{x,y| \hat{\rho} | x,y}.
\end{align}
The role of the off-diagonal part becomes evident when we compare
$P(x,x)$ and $P(x,-x)$ calculated for both $\rho_\infty$ and $\rho_\infty^\text{d}$
as illustrated in Fig.~\ref{fig:diagonal}.
While they are different when evaluated for the actual state $\rho_\infty$,
they are equal when evaluated for $\rho_\infty^\text{d}$
in which the off-diagonal part is absent.
In other words, the joint probability distribution shows a richer variety of behavior due to $\rho_\infty^\text{off-d}$
that cannot be captured by $\rho_\infty^\text{d}$ alone.

We point out a close analogy of our argument to the so-called decoherence-free subspace (DFS)~\cite{benatti2003}.
The simplest example of the DFS can be illustrated by a pair of qubits that are affected by random phase kicks~\cite{Palma1996}:
$(\ket{0}+\ee^{\ii \phi_1}\ket{1})(\ket{0}+\ee^{\ii \phi_2}\ket{1})$.
If the random phases $\phi_1$ and $\phi_2$ are independent,
averaging over them leads to the completely mixed state.
However, if they are correlated, for example, as $\phi_2=-\phi_1$,
quantum coherence remains nonvanishing, even after averaging
within a subspace spanned by $\ket{00}$ and $\ket{11}$.
In our model, the long-time average
corresponds to the phase averaging for each energy eigenstate.
In this process, there are degenerate subspaces~\eqref{eq:degen_sub},
where decoherence does not occur and quantum coherence is preserved.

\section{Statistical-Mechanical Description of entanglement pre-thermalization}\label{sec:ensembles}
So far, we have seen that our model captures the key features of EP.
In this section, we discuss the statistical-mechanical description of EP.
In Sec.~\ref{subsec:conventional},
we consider the canonical(-like) ensemble
by incorporating conserved quantities of each individual subsystem.
In Sec.~\ref{subsec:nonlocal},
we find an exact ensemble description by considering
nonlocal conserved quantities that nontrivially act on the entire system.

\subsection{Local conserved quantities}\label{subsec:conventional}
Here we argue that conventional statistical ensembles cannot describe $\rho_\infty$ well
as long as we take into account only conserved quantities that involve each individual subsystem.

The canonical ensemble
$\rho_{\rm can} \propto \ee^{-\beta \hat{H}_{\alpha=0}}= \sum_{m,n}\ee^{-\beta(m+n)}\ket{m,n}\bra{m,n}$,
obtained by maximizing the entropy under the constraint of the total energy $\hat{N}_1+\hat{N}_2$, is not sufficient
because it does not reproduce the basic property of $\rho_\infty$ that $\ket{m,n}$ is populated only when $m+n$ is even due to parity symmetry.
This constraint can be incorporated if we consider an extra conserved quantity $(-1)^{\hat{N}_1-\hat{N}_2}$,
which is unity for $\rho_\infty$.
The canonical ensemble with the constraint of $(-1)^{\hat{N}_1-\hat{N}_2}=1$ reads
\begin{align}\label{eq:can_even}
\rho_{\rm can}' = \sum_{N=0}^\infty \frac{\ee^{-2\beta' N}}{Z'} P_{2N},
\end{align}
where $P_{2N}=\sum_{m=0}^{2N}\ket{m,2N-m}\bra{m,2N-m}$
denotes the projection operator on the subspace with the total energy $2N$
and $Z'=\sum_{N=0}^\infty (2N+1) \ee^{-2\beta' N}$.
However, 
this still fails to approximate $\rho_\infty$ because $\rho_\text{can}'$ does not contain
any off-diagonal elements in the eigenenergy basis.

This difficulty cannot be overcome even if we consider extra local conserved quantities.
For instance, by considering any moment of the total energy $\hat{H}_{\alpha=0}^m$ $(m=2,3,\dots)$,
we obtain
\begin{align}\label{eq:can_weight}
\rho_\text{can}'' = \sum_{N=0}^\infty q_N P_{2N},
\end{align}
where $q_N$ is the actual weight on each degenerate subspace.
Although this ensemble contains many parameters,
it cannot describe EP because there is no off-diagonal element.
The ensemble~\eqref{eq:can_weight}
cannot approximate the long-time average~\eqref{eq:lta_rho} well
because it leads to $P(x,x)=P(x,-x)$ which does not hold in the actual long-time average.
This difficulty persists even if we take into account each weight $\braket{m,n| \rho | m,n}$
because it leads to $\rho_\infty^\text{d}$,
where no off-diagonal contribution is present.

\subsection{Nonlocal conserved quantities}\label{subsec:nonlocal}
In this subsection,
we show that, by taking into account nonlocal conserved quantities
in line with the rule of the generalized Gibbs ensemble,
we can construct an ensemble describing the long-time average exactly.

To this end, we work in the $a_\pm$ basis rather than the $a_{1,2}$ basis.
We note that nonlocal operators $\hat{N}_\pm \equiv a_\pm^\dag a_\pm$ are conserved: $[H_{\alpha=0},N_\pm]=0$
as seen from Eq.~\eqref{eq:ham0}.
By using these conserved quantities, we define a canonical-like ensemble
\begin{align}\label{eq:rho_nl0}
\rho_\text{NL}= \frac{1}{Z_{\rm NL}} \ee^{ -\sum_{\sigma=\pm} \beta_\sigma \hat{N}_\sigma },
\end{align}
where
$Z_{\rm NL}=\text{tr}\, \ee^{ -\sum_{\sigma=\pm} \beta_\sigma \hat{N}_\sigma }$,
and the parameters $\beta_\pm$ are determined by the conditions
 $\text{tr}(\rho_\text{NL} \hat{N}_+)=\braket{\psi(0)|\hat{N}_+|\psi(0)}=0$
and $\text{tr}(\rho_\text{NL} \hat{N}_-)=\braket{\psi(0)|\hat{N}_-|\psi(0)}=(\sinh r)^2$
that can be solved, giving $\beta_+=0$ and $\beta_-=2\ln\coth r$.
Then Eq.~\eqref{eq:rho_nl0} reduces to
\begin{align}\label{eq:nl1}
\rho_\text{NL}=(1-\ee^{-\beta_-})\sum_{N=0}^\infty \ee^{-\beta_- N}\kett{0,N}\!\braa{0,N}.
\end{align}

We can improve the ensemble by adding yet another conserved quantity $P_-\equiv (-1)^{N_-}$
as we have done in improving the ensemble from $\rho_\text{can}$ to $\rho_\text{can}'$.
The improved ensemble is given by
\begin{align}\label{eq:rho_nl}
\rho_{\rm NL}' = \frac{1}{Z_{\rm NL}'} \ee^{ -\sum_{\sigma=\pm} \beta_\sigma' \hat{N}_\sigma -\gamma P_-},
\end{align}
where $Z_{\rm NL}'=\text{tr}\, \ee^{ -\sum_{\sigma=\pm} \beta_\sigma' \hat{N}_\sigma -\gamma P_-}$
and the parameters $\beta_\pm'$ and $\gamma$ are determined by specifying 
the expectation values of $\hat{N}_\pm$ and $P_-\equiv (-1)^{N_-}$ in the initial state.
Concretely speaking,
our initial state~\eqref{eq:GSalpha}
satisfies $\braket{\psi(0)|N_+|\psi(0)}=0$,  $\braket{\psi(0)|N_-|\psi(0)}=(\sinh r)^2$ and $\braket{\psi(0)|P_-|\psi(0)}=1$,
which imply
\begin{align}\label{eq:nl2}
\rho_\text{NL}'=(1-\ee^{-2\beta_-'})\sum_{N=0}^\infty \ee^{-2\beta_-' N}\kett{0,2N}\!\braa{0,2N},
\end{align}
where $\beta_-'=\frac{1}{2}\ln(1+\frac{2}{\sinh^2 r})$
and the symbol $\kett{\cdots}$ denotes the state in the $a_\pm$ basis (see Eq.~\eqref{eq:pm_basis}).
We note that $\rho_\text{NL}'$ is similar in form to $\rho_\infty=\sum_{N=0}^\infty q_N \kett{0,2N}\! \braa{0,2N}$.

Furthermore,
by taking more conserved quantities $\{\hat{N}_-^m\}_{m=2}^\infty$
in line with the rule of the generalized Gibbs ensemble,
we obtain the exact final-state distribution of the system:
\begin{align}
\rho_{\rm NL}''=
\frac{1}{Z_{\rm NL}''} \ee^{-\sum_{m=1}^\infty \beta_-^{(m)} \hat{N}_-^m -\beta_+ \hat{N}_+ -\gamma P_-}
=\rho_\infty.
\end{align}
In deriving the last equality,
we have used the solution $\{\beta_-^{(m)}\}_{m=1}^\infty$
to the set of equations $\text{tr}(\rho_{\rm NL}\hat{N}_-^m )=\braket{\gs|\hat{N}_-^m |\gs}$ ($m=1,2,\dots$).
This systematic improvement of $\rho_{\rm NL}$ to the exact result can be successfully made because
$\rho_\infty$ becomes diagonal in the nonlocal $a_\pm$ basis (see Eq.~\eqref{eq:lta_rho}),
while, in the local $a_{1,2}$ basis, $\rho_\infty$ involves the off-diagonal part $\rho_\infty^\text{off-d}$,
which can be incorporated by neither the canonical ensemble nor its generalization.

\begin{figure}[b]
\begin{center}
\includegraphics[width=8.5cm]{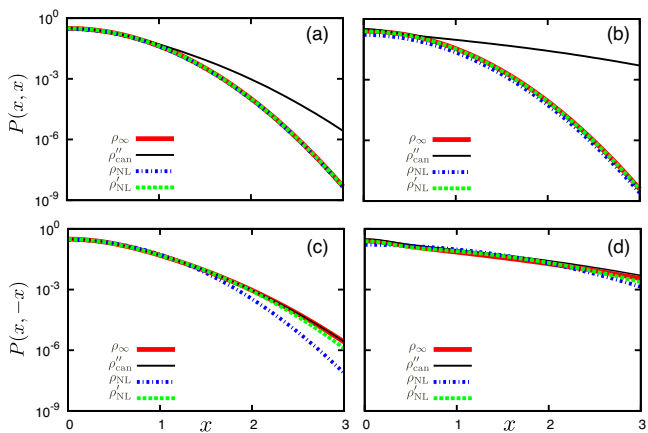}
\caption{(Color Online)
Distribution $P(x,y)$ with $x=y$ (upper two panels)
and $x=-y$ (lower two panels) calculated for
the long-time average~\eqref{eq:lta_rho} (thick solid),
the canonical-like ensemble~\eqref{eq:can_weight} (thin solid),
and the canonical-like ensemble with nonlocal conserved quantities~\eqref{eq:nl1} (dashed) and \eqref{eq:nl2} (dash-dotted)
plotted for the quench magnitude of $r=0.3$ (left two panels) and 1 (right two panels).
}
\label{fig:nls}
\end{center}
\end{figure}

Figure~\ref{fig:nls} shows how these ensembles give increasingly better results for the joint probability distribution.
It is clearly seen that $\rho_\text{NL}'$ gives a better prediction 
than $\rho_\text{can}''$
even though $\rho_\text{NL}'$ has much less fitting parameters.
This observation implies that an appropriate choice of basis 
is more important than the number of parameters.
Figure~\ref{fig:nls} also shows that $\rho_\text{NL}'$ gives even better predictions
than $\rho_\text{NL}$ as expected.
As we have shown,
by going further to $\rho_\text{NL}''$,
we obtain the exact distribution.

Finally, we discuss how we could know, in more general situations, the best set of conserved quantities
like $\hat{N}_\pm$ rather than $\hat{N}_i$ $(i=1,2)$.
This is possible if we are given the expectation values of the lowest-order correlation in, say,
the $a_i$ ($i=1,2$) basis:
\begin{align}
A_{ij} = \braket{\psi(0)| a_i^\dag a_j | \psi(0)}=
\frac{(\sinh r)^2}{2}
\begin{pmatrix}
1 & -1\\
-1 & 1
\end{pmatrix}_{ij}.
\end{align}
The natural candidate for the statistical ensemble describing $\rho_\infty$
is the canonical-like ensemble which is obtained by taking into account all the low-order conserved quantities.
We note that there are four linearly independent conserved quantities
since $Q\equiv \sum_{i,j}a_i^\dag Q_{ij} a_j $ for any $2\times2$ Hermitian matix commutes with $\hat{H}_{\alpha=0}$.
Thus we can take $\sigma^\gamma = \sum_{i,j}a_i^\dag \sigma^\gamma_{ij} a_j$ $(\gamma=0,1,2,3)$,
where $\sigma^0_{ij}$ is the $2\times2$ identity matrix and the other three are the Pauli matrices.
Then, by introducing the Lagrange multipliers $\lambda_\gamma$ $(\gamma=0,1,2,3)$,
our lowest-order ensemble is given by
\begin{align}\label{eq:lo1}
\rho_\text{LO}=\frac{1}{Z_\text{LO}}\exp\left( -\sum_{\gamma=0}^3 \lambda_\gamma \sigma^\gamma\right),
\end{align}
where the Lagrange multipliers are determined by
\begin{align}
\text{tr} (\rho_\text{LO} a_i^\dag a_j) = A_{ij}.
\end{align}
By solving these equations, one can show $\lambda_2=\lambda_3=0$,
which implies that Eq.~\eqref{eq:lo1} reduces to Eq.~\eqref{eq:rho_nl0} with $\beta_\pm = \lambda_0\pm \lambda_1$.
Thus, we obtain
\begin{align}
\rho_\text{LO} = \rho_\text{NL}
\end{align}
without assuming a priori knowledge of the proper basis $a_\pm$.
We note that this is also justified through the diagonalization of $A_{ij}$,
which amounts to the change of basis from $a_{1,2}$ to $a_\pm$.

\section{Entanglement Pre-thermalization Plateau}\label{sec:plateau}
EP originates from the degeneracy in the energy spectrum due to some symmetry of the system.
However, 
the symmetry may only be approximate in reality.
In this case, EP is observed as a transient phenomenon which eventually relaxes to a thermal equilibrium state.
In this section, we demonstrate this by slightly breaking the symmetry between the two systems.

Suppose that the Hamiltonian after the coherent splitting
is given, instead of Eq.~\eqref{eq:ham}, by
\begin{align}\label{eq:ham_epsilon}
\hat{H}_{\alpha=0}^\epsilon= (1+\epsilon) a_1^\dag a_1 + a_2^\dag a_2
\end{align}
and that the Hamiltonian before the splitting is given by Eq.~\eqref{eq:ham_epsilon} with $\epsilon=0$.
Here a small real number $\epsilon(\neq0)$ quantifies the asymmetry between two potentials after the splitting.
We assume that $\epsilon$ is an irrational number to ensure no degeneracy in the energy spectrum.

Under these conditions,
only the diagonal part contributes to the long-time average.
The time-evolved state is given, instead of Eq.~\eqref{eq:timet}, by
\begin{align}
\ket{\Psi_\epsilon(t)}&=\sum_{N=0}^\infty\sum_{m=0}^{2N} (-1)^N\sqrt{q_N}c_{m,2N-m}\notag\\
&\qquad\qquad\times
\ee^{-\ii(2N+m\epsilon)t}\ket{m,2N-m},
\end{align}
and thus the long-time average is obtained as
\begin{align}
\rho_\infty^\epsilon=\overline{\ket{\Psi_\epsilon(t)}\bra{\Psi_\epsilon(t)}}=\rho_\infty^\text{d},
\end{align}
where $\rho_\infty^\text{d}$ is given in Eq.~\eqref{eq:diagonal}.
This implies that the system eventually loses the characteristics of EP
and becomes describable by the standard statistical ensembles discussed above.

We now investigate the transient behavior
by examining the time evolution of an observable
\begin{align}
\hat{O}=(\hat{x}_1-\hat{x}_2)^2.
\end{align}
The expectation value of this quantity for any state $\hat{\rho}$ is related to $P(x,y)$ defined in Eq.~\eqref{eq:joint}
as $\text{tr}(\hat{\rho} \hat{O})=\int \dd x \dd y P(x,y) (x-y)^2$.
In discussing the time evolution,
we consider the following time-averaged quantity
\begin{align}\label{eq:OT}
O_T \equiv \int_0^T \frac{\dd t}{T} \braket{\Psi_\epsilon(t)| \hat{O} | \Psi_\epsilon(t)}
\end{align}
rather than the time-dependent expectation value
$\braket{\Psi_\epsilon(t)| \hat{O} | \Psi_\epsilon(t)}$
since the latter keeps oscillating due to numerous equal energy-level spacings in the energy spectrum.
In more realistic systems, this oscillation is known to be exponentially suppressed
with increasing the system size~\cite{Reimann2008,Short2011}.

The time-averaged expectation value~\eqref{eq:OT} is exactly calculated to give
\begin{widetext}
\begin{align}\label{eq:OT_analytic}
O_T = 1+\sinh^2 r -\frac{\sinh(2r)}{8}\frac{\sin(2T)}{T}
-\frac{\sinh(2r)}{8}\left\{
\frac{\sin[2(1+\epsilon)T]}{(1+\epsilon)T}
+4\frac{\sin[(2+\epsilon)T]}{(2+\epsilon)T}
\right\}
+\sinh^2 r \frac{\sin(\epsilon T)}{\epsilon T}.
\end{align}
\end{widetext}
One can easily confirm that, in the limit of $\epsilon\to0$, Eq.~\eqref{eq:OT_analytic}
reduces to
\begin{align}
O_T=1+2\sinh^2r-\sinh(2r)\frac{\sin(2T)}{2T} \quad (\epsilon=0).
\end{align}
Thus, in the absence of the symmetry-breaking perturbation $\epsilon$,
$O_T$ converges to the EP value of $1+2\sinh^2 r$
which is given by $\text{tr}(\hrho_\infty \hO)$,
whereas, in its presence, it approaches a different value $1+\sinh^2 r$,
which is given by $\text{tr}(\rho^\text{d}_\infty O)$.

\begin{figure}
\begin{center}
\includegraphics[width=8.5cm]{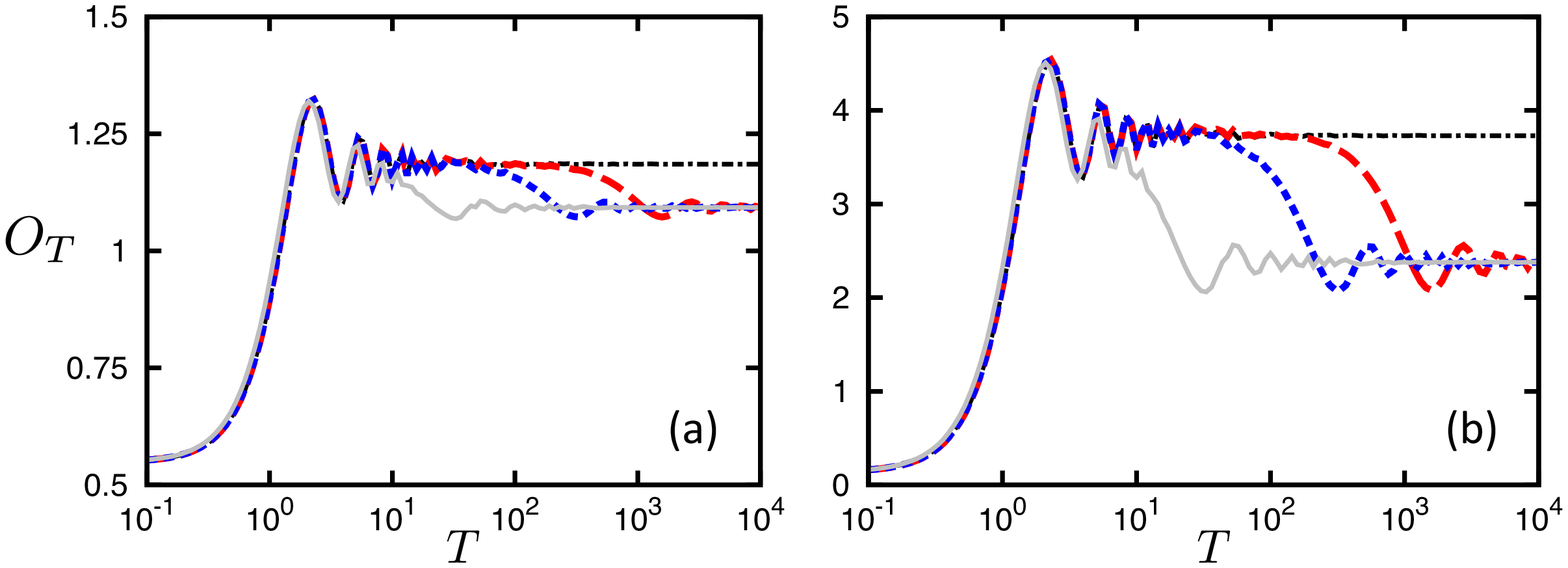}
\caption{(Color Online)
Time averages for the observable defined in Eq.~\eqref{eq:OT}
for the quench magnitude $r=0.3$ (a) and $1$ (b),
where $O_T$ is plotted for four different values of $\epsilon=0$ (dash-dotted),
$2\sqrt{2}\times10^{-3}$ (dashed), $\sqrt{2}\times10^{-2}$ (dotted),
and $\sqrt{2}\times10^{-1}$ (solid).
The entanglement pre-thermalization plateau appears for sufficiently small $\epsilon$.
}
\label{fig:evolution}
\end{center}
\end{figure}

The EP plateau appears for sufficiently small $\epsilon$ as shown in Fig.~\ref{fig:evolution}.
For sufficiently small $\epsilon$ such as $2\sqrt{2}\times10^{-3}$ and $\sqrt{2}\times10^{-2}$,
$O_T$ first approaches the EP value, stay there for a while,
and then converges to the value given by $\text{tr}(\rho^\text{d}_\infty O)$.
This is because
the symmetry-breaking perturbation
$\epsilon$ introduces yet another energy scale and, hence, another time scale $1/\epsilon$.
For $T\ll 1/\epsilon$, the dynamics is well approximated by the one with $\epsilon=0$,
whereas it is modified after $T=O(\epsilon^{-1})$
as evident from Eq.~\eqref{eq:OT_analytic}.
Thus, the duration of the EP plateau
is of the order of $1/\epsilon$.

The EP plateau disappears if the symmetry-breaking perturbation $\epsilon$
becomes comparable with the energy-level spacings of the unperturbed system.
As illustrated in Fig.~\ref{fig:evolution},
for $\epsilon=\sqrt{2}\times10^{-1}$,
the EP plateau disappears and $O_T$ directly converges to $\text{tr}(\rho^\text{d}_\infty O)$.
This is because the time scale $1/\epsilon$ is comparable with
the intrinsic scale, which has been set to unity in our discussion.

We expect that such a plateau-like behavior is commonly seen
for generic systems that exhibit EP.
EP derives from some symmetry that ensures the degeneracy in the energy spectrum
and thus preserves quantum entanglement.
On the other hand, in experimentally realistic systems,
there may be symmetry-breaking perturbations,
and EP appears only as a plateau.
The duration of EP is expected to be $O(\epsilon^{-1})$,
where $\epsilon$ represents the magnitude of the perturbation,
because the energy splitting of the degenerate eigenstates is $O(\epsilon)$
and this does not affect the dynamics in that short-time scale.

\section{Conclusions}\label{sec:conclusions}
We have analyzed EP by using a simple model consisting of two harmonic oscillators
and shown that this simple model involves the characteristics of EP proposed in Ref.~\cite{Kaminishi2015}.
Namely, while each individual subsystem is asymptotically thermal in the long-time limit,
the total system is not due to quantum entanglement present in the initial state.
We have taken a step further to identify the statistical ensemble that describes the nonthermal total system.
We have shown that, if we know low-order correlations between the subsystems,
we can construct a GGE-like ensemble, which exactly describes the total system.
Finally, we have discussed the effect of small symmetry-breaking perturbations.
If the perturbation is sufficiently small, EP is rather observed as a plateau
in the time evolution of an observable.

Since the study of EP is still in its infancy,
there are many open questions to be addressed.
First of all, more examples are needed to
gain a comprehensive understanding of EP.
Second,
it is important to study nonintegrable setups,
since only integrable systems have been studied so far.
The generalization to a mixed state in an open system is of interest to understand the range of applicability of EP.
Our finding of a close relationship between EP and the DFS
might give a useful
insight into when quantum entanglement survives in the thermodynamic situation.
Of course, experimental investigations are of great interest.

\section*{Acknowledgements}
This work was supported by
JSPS KAKENHI Grant Numbers JP16H06718, JP15K17718, JP16J03140, JP26287088, 
a Grant-in-Aid for Scientific Research on Innovative Areas ``Topological Materials Science'' (KAKENHI Grant No. JP15H05855), 
and the Photon Frontier Network Program from MEXT of Japan.
T.N.I. acknowledges the JSPS for the postdoctoral fellowship for research abroad.

\bibliography{draft_arXiv}

\end{document}